\documentclass[final,copyright,creativecommons]{eptcs}
\providecommand{\event}{SCAV 2018}

\usepackage{breakurl,underscore} 
\usepackage{amsmath,amssymb,bm,upgreek,theorem}
\usepackage{float,epsfig,pst-node,subfig,wrapfig,xcolor}
\usepackage{pstricks,auto-pst-pdf}

\psset{unit=1mm,framesep=10pt,fillstyle=none}  
\degrees[360]                                  
\psset{linewidth=0.9pt}                        
\providecommand{\mel}{\psset{linewidth=0.5pt}} 
\providecommand{\stl}{\psset{linewidth=0.9pt}} 

\newlength\figureheight	
\newlength\figurewidth	

\title{On the Application of ISO 26262 in Control Design for Automated Vehicles
}

\author{Georg Schildbach
\institute{Institute for Electrical Engineering in Medicine, University of Luebeck}
\email{georg.schildbach@uni-luebeck.de}
}

\def\titlerunning{ISO 26262 for Control Design for AVs}
\def\authorrunning{G.\ Schildbach}

\begin{document}

\maketitle
\thispagestyle{empty}
\pagestyle{empty}


\begin{abstract}
Research on automated vehicles has experienced an explosive growth over the past decade. A main obstacle to their practical realization, however, is a convincing safety concept. This question becomes ever more important as more sophisticated algorithms are used and the vehicle automation level increases. The field of functional safety offers a systematic approach to identify possible sources of risk and to improve the safety of a vehicle. It is based on practical experience across the aerospace, process and other industries over multiple decades. This experience is compiled in the functional safety standard for the automotive domain, ISO 26262, which is widely adopted throughout the automotive industry. However, its applicability and relevance for highly automated vehicles is subject to a controversial debate. This paper takes a critical look at the discussion and summarizes the main steps of ISO 26262 for a safe control design for automated vehicles.
\end{abstract}


\section{INTRODUCTION}\label{Sec:Intro}

\emph{Automated vehicles} (AVs), that is vehicles with automation level 3 or higher \cite{VDAMag:2015}, are the subject of enormous research interest. A lot of their foundational technology has now become mature. Many experts consider AVs to be at the verge of market introduction, with the potential to spark a revolution in the transportation sector. A major argument in favor of AVs is that they improve traffic safety and reduce fatal accidents. The rationale is that sophisticated algorithms running on ever more powerful hardware can be more reliable, vigilant, and knowledgable than the average human driver. Yet the delicate question on how to certify the safety of AVs remains wide open \cite{KoopWag:2016}. 

This paper is mainly concerned with the design of controllers for AVs. Besides the classical robustness properties of a controller, another important factor for safety is the development process. In control design, idealized assumptions are typically made that are not always satisfied in practical operation. Moreover, the controller logic is implemented on electronic hardware, whose computations may be wrong or which may even fail completely. The controller software is developed and programmed by humans, working together in teams, using uncertified software tools, and is therefore subject to conceptual and implementation errors.

All of these errors entail a risk that the overall system fails in fulfilling some or all of its intended functionalities. The identification of risk sources and their severity, as well as possible measures for mitigation, cannot be an exact science. \emph{Functional safety} (FS) offers a systematic approach to this end.

Only recently, in 2011, ISO 26262 \cite{ISO26262:2011} has been introduced as the domain specific standard for FS, refining the more general IEC 61508 for the automotive industry. It is concerned with the safety processes that accompany the product development, known as the `safety lifecycle'. A \emph{controller} in the sense of ISO 26262 is understood, very generally, as a programmable software that runs on an embedded hardware platform, whose logic maps input information (e.g., from sensors) and control objectives to output commands (e.g., for actuators). This paper uses a narrower definition of \emph{controller}, in the sense of control theory, using concepts such as machine learning \cite{LefevreEtAl:2016,VallonEtAl:2017} or model predictive control \cite{KongEtAl:2015,Schildi:2016}.

\section{BACKGROUND}\label{Sec:Background}

\vspace*{-0.06cm}
\subsection{A Legal Perspective on ISO 26262}\label{Sec:Legal}

Besides the obvious ethical aspects of protecting human health and life, FS has important legal implications. As an industry standard, ISO 26262 is not an immediate law in the United States (U.S.) and the European Union (E.U.). The legal relevance of ISO 26262 is therefore indirect. Due to its recent introduction, there exist little to no known precedent legal cases involving ISO 26262. Moreover, existing law suits can be expected to settle out of court in most cases. Therefore the exact legal implications of ISO 26262 are subject to debate among experts \cite{Gollob:2007,Helmig:2014}.

ISO 26262 has been developed for safety critical systems in conventional vehicles, i.e., driven by humans. Examples are airbags, stability control, or an emergency brake assist. The legal territory of AVs is, by and large, uncharted. Without further clarification by legislators, regulators, or legal precedents, the limit on what is legally permitted or prohibited, in terms of automation technology, remains an educated guess \cite{Smith:2012}. It is likely, however, that autonomous vehicles are generally legal in the U.S.\ and the E.U., or can be expected to become legal in the near future \cite{Pillath:2016,Smith:2012}.

In the E.U., a vehicle manufacturer is liable for any damages caused by defects in his products \cite{Gollob:2007}, including AVs. This liability holds also \emph{without an actual fault} of the manufacturer. Some countries, like Germany, make an exception if the manufacturer can prove that it was not possible to avoid or realize the defect at the time of product delivery, based on the \emph{current state of technology}. Concerning FS, the current state of technology is deemed to be compiled, more or less completely, in ISO 26262 \cite{Helmig:2014}.

In the U.S., product liability is more complicated, due to a stronger case law and the dual existence of a federal and state court system. Similar to the E.U., the manufacturer can be liable for defects of his products \emph{without any own fault}, through implied warranty or strict liability in tort \cite{Gollob:2007}. A defence based on the \emph{current state of technology} is permitted only in some of the states. However, the case of the manufacturer is strongly supported if he can show that the efforts made for the safety of his product conform with, or even surpass, the best practices of the industry \cite{Wu:2016}.

\vspace*{-0.06cm}
\subsection{On the Scope of Functional Safety}\label{Sec:FunSafe}

According to ISO 26262, functional safety is the `absence of unreasonable risk due to hazards caused by malfunctioning behavior of e/e systems' \cite[part 1.51]{ISO26262:2011}. This definition lists four fundamental criteria for the scope of FS:

\vspace*{-0.1cm}
\begin{enumerate}
	\setlength\itemsep{1pt}
	\item ISO 26262 is concerned with \emph{e/e systems}, i.e., systems containing electrical and/or electronic components, such as sensors, actuators, or programmable controllers.
	\item A \emph{hazard} is a potential source of \emph{harm}, i.e., 'a physical injury or damage to the health of persons'. In particular, ISO 26262 is not concerned with damages to property, such as the car itself.
	\item \emph{Risk} refers to the `combination of the probability of occurrence of harm and the severity of that harm'. It is a basic principle of ISO 26262 that it is not practically possible to eliminate \emph{all} risk, but \emph{unreasonable} risk. A risk is unreasonable if it is 'judged to be unacceptable in a certain context according to valid societal moral concepts'. 
	\item \emph{Malfunctioning behavior} of an item comprises any 'failure or unintended behavior [...] with respect to the design intent'. Hence, FS only covers faults or errors in the item's hardware or software if they lead to a deviation from their intended behavior. Conversely, FS is not concerned with the safety of the \emph{intended behavior} of a system. To this end, a new standard for the \emph{safety of the intended function} (SOTIF) is under way, which will complement ISO 26262.
\end{enumerate}

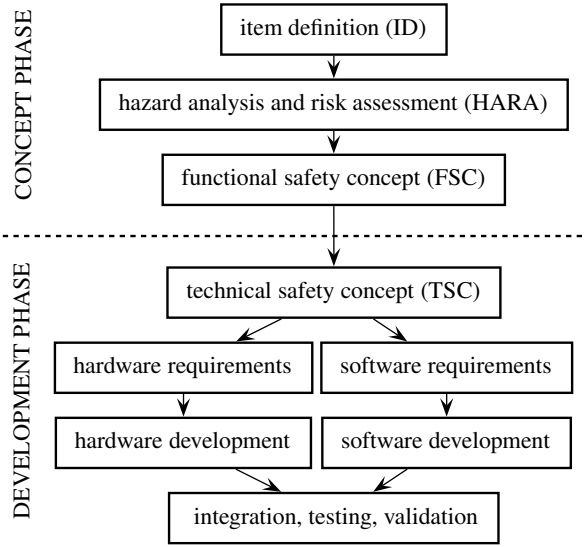
\begin{wrapfigure}{r}{0.48\textwidth}
	\begin{center}

		\psset{unit=1.0mm}
        \hspace*{0.3cm}
		\begin{pspicture}(0,0)(70,60)
		\footnotesize
		\stl
		\psset{arrowsize=5pt}
						
		\rput[bc](37,55){\rnode{ID}{\psframebox[framesep=5pt]{\phantom{p}\hspace*{-0.1cm}item
		definition (ID)\phantom{P}\hspace*{-0.1cm}}}}
		\rput[bc](37,45){\rnode{HARA}{\psframebox[framesep=5pt]{\phantom{p}\hspace*{-0.1cm}hazard
		analysis and risk assessment (HARA)\phantom{P}\hspace*{-0.1cm}}}}
		\rput[bc](37,35){\rnode{FSC}{\psframebox[framesep=5pt]{\phantom{p}\hspace*{-0.1cm}functional
		safety concept (FSC)\phantom{P}\hspace*{-0.1cm}}}}
		\rput[bc](37,20){\rnode{TSC}{\psframebox[framesep=5pt]{\phantom{p}\hspace*{-0.1cm}technical
		safety concept (TSC)\phantom{P}\hspace*{-0.1cm}}}}
		\rput[bc](17,10){\rnode{HR}{\psframebox[framesep=5pt]{\phantom{p}\hspace*{-0.1cm}hardware
		requirements\phantom{P}\hspace*{-0.1cm}}}}
		\rput[bc](52,10){\rnode{SR}{\psframebox[framesep=5pt]{\phantom{p}\hspace*{-0.1cm}software
		requirements\phantom{P}\hspace*{-0.1cm}}}}
		\rput[bc](17,0){\rnode{HD}{\psframebox[framesep=5pt]{\phantom{p}\hspace*{-0.1cm}hardware
		development\phantom{P}\hspace*{-0.1cm}}}}
		\rput[bc](52,0){\rnode{SD}{\psframebox[framesep=5pt]{\phantom{p}\hspace*{-0.1cm}software
		development\phantom{P}\hspace*{-0.1cm}}}}
		\rput[bc](37,-10){\rnode{ITV}{\psframebox[framesep=5pt]{\phantom{p}\hspace*{-0.1cm}
		integration, testing, validation\phantom{P}\hspace*{-0.1cm}}}}
		\mel
		\ncline[nodesep=0pt]{->}{ID}{HARA}
		\ncline[nodesep=0pt]{->}{HARA}{FSC}
		\ncline[nodesep=0pt]{->}{FSC}{TSC}
		\ncline[nodesep=0pt]{->}{TSC}{HR}
		\ncline[nodesep=0pt]{->}{TSC}{SR}
		\ncline[nodesep=0pt]{->}{HR}{HD}
		\ncline[nodesep=0pt]{->}{SR}{SD}
		\ncline[nodesep=0pt]{->}{HD}{ITV}
		\ncline[nodesep=0pt]{->}{SD}{ITV}
		\stl
		\psline[linestyle=dashed,dash=2pt 2pt](-7,27.5)(70,27.5)
		\rotateleft{\rput[bc](45,4){\psframebox[linestyle=none]{\phantom{p}\hspace*{-0.1cm}CONCEPT PHASE\phantom{P}\hspace*{-0.1cm}}}}
		\rotateleft{\rput[bc](7,5){\psframebox[linestyle=none]{\phantom{p}\hspace*{-0.1cm}DEVELOPMENT PHASE\phantom{P}\hspace*{-0.1cm}}}}
		\end{pspicture}
		\psset{unit=1.0mm}
	\end{center}
	\vspace*{1.0cm}
	\caption{Main steps of the ISO 26262 development process.\label{Fig:ISOOverview}}
\end{wrapfigure}

FS is thus but one aspect of the overall safety of an AV. Other aspects include the robustness oagainst environmental influences (e.g., variations in temperature, humidity, chemicals, radiation) and the design of the human machine interface. The latter must be intuitive and not allow for unintentional, unsafe operations (e.g., accidental activation of a lever with a knee). Last but not least, security against malicious attacks is becoming an increasingly important factor for safety in a connected, digital world.

The FS development process prescribed by ISO 26262 mirrors the classical V-model for automotive electronics \cite{Sommerville:2016}. Fig.\ \ref{Fig:ISOOverview} depicts the major steps. The process starts on the top left of the V with a definition of the functionality. Then the safety requirements are deduced and gradually refined, while moving down the left side of the V. At the lower tip, the actual hardware and software are developed. Moving up on the right side of the V, safety requirements from the left side are verified. There are additional processes, such as management of the FS process and supervision during the production and operation phase.

\subsection{A Technical Perspective on ISO 26262}

With its introduction in 2011, ISO 26262 is still fairly recent. It has been widely accepted and adopted within the automotive industry. However, its roots lie in traditional automotive control systems and it has not been designed with view on the new challenges of AVs. Thus ISO 26262 is also subject to a technical controversy. The main points of criticism, in with a particular perspective on control design, will be listed throughout the next sections, e.g., the one that has already appeared above.\\

\noindent\fbox{\begin{minipage}{\dimexpr\textwidth-2\fboxsep-3\fboxrule\relax}
\textbf{(A) SOTIF and Security:} ISO 26262 has a narrow perspective on safety, not considering the aspects of SOTIF and security. For traditional automotive control systems, such as airbags or stability control, these are either trivial or of minor relevance. For AVs, however, they represent a major challenge.
\end{minipage}}

\section{CONTROLLER DESIGN: CONCEPT PHASE}\label{Sec:ConPhase}

It is beyond the scope of this paper to describe all processes of ISO 26262 in detail \cite{Hillenbrand:2011}. Instead, the goal is to discuss the main steps with respect to control design. This section covers the \emph{concept phase} \cite[part 3]{ISO26262:2011}, which is concerned with the system from a \emph{functional perspective}. Details about the technical implementation are not relevant at this stage, such as the software architecture or hardware platform.

\subsection{Item Definition (ID)}

The safety process of ISO 26262 starts with the definition of the function, more generally called an \emph{item} \cite[part 3.5]{ISO26262:2011}. The ID defines the item from a non-technical, functional perspective at the vehicle level. This includes its nominal behavior, including the system limits, as well as its interfaces and the interactions with the immediate environment. In contrast, the physical realization of the item -- i.e., in hardware and software -- is called the \emph{system}. The item must be such that its system contains (at least) one sensor, one controller, and one actuator. The safety concept must be developed for the entire item, in this case the AV, of which the controller is an integral part.\\

\noindent\fbox{\begin{minipage}{\dimexpr\textwidth-2\fboxsep-3\fboxrule\relax}
\textbf{(B) Item definition:} AVs have to fulfil numerous complex tasks under diverse environmental conditions (roads, weather, animals, other drivers, equipment failures, etc.). Hence, even more than usual \cite{Sommerville:2016}, generating a set of valid, consistent, complete, realistic, and verifiable requirements becomes a major challenge. The number of requirements explodes, and may even become infinite.
\end{minipage}}

\subsection{Hazard Analysis and Risk Assessment (HARA)}

The first step of HARA \cite[part 3.7]{ISO26262:2011} is \emph{hazard identification}. Based on the ID, possible malfunctions of the item are listed that may lead to hazards. Typical techniques to this end are either inductive (e.g., brainstorming, FMEA) or empirical (e.g., quality history, field studies).

The second step of HARA is to perform a \emph{hazard analysis}. The item's malfunctions are considered in different operating scenarios, where they may lead to an \emph{hazardous event} (HE). If so, the \emph{exposure} to that operating scenario is recorded and classified into the categories E0 (`almost never') to E4 (`very frequently'). For a wide range of scenarios, suitable values for the exposure can be extracted from standard tables, e.g., by the German Association of the Automotive Industry (VDA).

The third step consists of a \emph{risk assessment}. In addition to the exposure, each HE is assessed according to its severity and its controllability. \emph{Severity} measures the harm that the event can potentially entail and ranges from S0 (`no injuries') to S3 (`fatal injuries'). \emph{Controllability} means the ability of an average driver, who is in suitable condition and has a standard driver's training, to avoid the harm. It is classified between C0 (`controllable in general') to C3 (`difficult or impossible to control').

According to its E,S,C-values, each HE is assigned an \emph{automotive safety integrity level} (ASIL). It is based on the sum of E\,+\,S\,+\,C, if all of these summands are non-zero. HEs with the highest possible sum of 10 (E4\,+\,S3\,+\,C3) have the strongest safety requirements, and are assigned an 'ASIL D'. Decreasing safety requirements follow for HEs with sum values of 9, 8, and 7, which are assigned an 'ASIL C', 'ASIL B', and 'ASIL A', respectively. Sum values of 6 and below, or if at least one of the summands is zero, do not receive an ASIL. They are assigned a simple 'quality management' (QM), which means that they do not require any particular safety measures beyond the baseline engineering practices.

For each HE, one (or more) \emph{safety goal(s)} (SGs) are derived. The SG is usually to prevent the malfunction that causes the HE, to mitigate its severity and/or to improve its controllability by the driver.\\

\noindent\fbox{\begin{minipage}{\dimexpr\textwidth-2\fboxsep-3\fboxrule\relax}
\textbf{(C) Safety validation:} Many AV algorithms are strongly based on machine learning, e.g., deep learning or even deep driving. Other algorithms produce non-deterministic results. For control algorithms, in particular, SGs may be dependent on the tuning of a large parameter set. For these algorithms, it is difficult to enforce or verify a-priori set SGs.
\end{minipage}}\\

\vspace*{0.3cm}
\noindent\fbox{\begin{minipage}{\dimexpr\textwidth-2\fboxsep-3\fboxrule\relax}
\textbf{(D) Safety levels:} For an AV, with an automation level 3 or higher \cite{VDAMag:2015}, the controllability value will generally rise to C3, as the driver may be incapable to exert any control over the vehicle (e.g., due to a disability, blindness, no driver training, etc.) \cite{KoopWag:2016}. Hence its SGs are generally assigned high ASILs.
\end{minipage}}

\subsection{Functional Safety Concept (FSC)}\label{Sec:FSC}

For each SG of the HARA, the FSC \cite[part 3.8]{ISO26262:2011} defines appropriate safety mechanisms or safety measures, called \emph{functional safety requirements} (FSRs). A SG may entail multiple FSRs and one FSR may be shared by multiple SGs. Each FSR inherits the highest ASIL of all associated SGs. \emph{ASIL inheritance} is a common thread running through ISO 26262. It means that the child requirement receives the maximum ASIL of all of its parents. 

The FSRs may introduce additional elements to the item's functional architecture. FSRs may include, for instance, a supervisor for fault detection, a mechanism for fault tolerance, the transition to a safe state, a driver warning, or a redundancy of components.

Another important concept of ISO 26262 is the \emph{ASIL decomposition} \cite[part 9.5]{ISO26262:2011}. Its effect is to lower the ASILs along the inheritance chain. An ASIL decomposition applies if two (or more) child requirements or architectural elements are \emph{redundant} towards their parent requirement or architectural element (i.e., they are both able to satisfy their parent). Moreover, they must be \emph{sufficiently independent}.\\



\noindent\fbox{\begin{minipage}{\dimexpr\textwidth-2\fboxsep-3\fboxrule\relax}
\textbf{(E) Fail-silence:} ISO 26262 is designed primarily with fail-silent functions in mind. Thus it is a tacit assumption that the vehicle reaches a \emph{safe state} when a fault is detected and the function is switched off. This does not hold for AVs, however, which have to remain fail-operational at all times.
\end{minipage}}

\section{CONTROLLER DESIGN: DEVELOPMENT PHASE}\label{Sec:DevPhase}

In the development phase, the FSC is merged with the \emph{technical details} of the specific implementation. The technical details are documented in the \emph{system design specification} (SDS), describing the hardware and software architecture, all interfaces, and the non-safety requirements \cite[part 4.7]{ISO26262:2011}. 

The \emph{technical safety concept} (TSC) \cite[part 4.6]{ISO26262:2011} refines the FSRs to \emph{technical safety requirements} (TSRs). A TSR inherits the ASIL from its parent FSRs (as described above). In the TSC, all TSRs are mapped to their corresponding software and hardware components, as specified in the SDS.

\subsection{Hardware Design}

For hardware safety analyses, computation results are considered as probabilistic \cite{Efody:2018}, with faults occurring infrequently at random. The nature of these faults can be one-time (e.g., a gate giving a wrong value in one cycle) or permanent (e.g., a gate getting stuck at a fixed value). Their probability is measured as \emph{failures in time} (fit), with 1 fit corresponding to 1 fault occurring in $10^{9}$ hours. The main concern of FS is to limit the general occurrence faults and/or prevent them from compromising the SGs.

Safety for hardware design splits into two parts. The first us about the \emph{robustness of the hardware architecture} against critical faults \cite[part 5.8]{ISO26262:2011}. Depending on whether there exists a redundancy mechanism or not, they fall under the \emph{latent fault metric} (LFM) or the \emph{single-point fault metric} (SPFM) \cite[part 5.C]{ISO26262:2011}. Moreover, a \emph{diagnostic coverage} (DC) is computed \cite[part 5.D]{ISO26262:2011}. The second part is concerned with the violation of specific safety goals based on a \emph{probabilistic metric for random hardware failures} (PMHF) \cite[part 5.9]{ISO26262:2011}. Each hardware component inherits an ASIL from its associated TSRs. Permissible target values for PMHF are 10 fit for ASIL D, 100 fit for ASIL C and ASIL B components. 

Many items, however, require more advanced hardware architectures, such as ECUs with microprocessors. In this case a detailed analysis down to the level of logical gates becomes impractical and faults are considered on a more abstract level. To this end, specific ECUs have been designed that comply with ISO 26262 as a \emph{safety element out of context} (SEooC). This means that a safety manual specifies the PMHF, for example, as 3 fit, subsuming all kinds of faults in the ECU, which may or may not violate a SG. The remaining PMHF budget can thus be distributed over the other hardware components based on a more detailed analysis \cite{JohKar:2015}.\\

\noindent\fbox{\begin{minipage}{\dimexpr\textwidth-2\fboxsep-3\fboxrule\relax}
\textbf{(F) Hardware complexity:} The hardware needed for AVs is by several magnitudes more powerful than that of traditional automotive control systems. Hence many of the detailed techniques of ISO 26262 for safety analysis become very burdensome or even impossible to implement.
\end{minipage}}

\subsection{Software Development}

Analogous to hardware, a software component inherits the highest ASIL from the TSRs that it implements. The development process for safe software, however, is fundamentally different from that of hardware. One reason is that software does not fail randomly like hardware \cite{ThomasEtAl:2012}. Errors in software are usually systematic and the result of erroneous requirements specifications or coding bugs \cite{Sommerville:2016}. 

In contrast to hardware, software cannot achieve reliable protection by redundancy. As shown in \cite{KnightLev:1986,KnightLev:1990}, multiple versions of the same software produced by independent teams are likely to contain the same errors. Finally, software cannot be tested as exhaustively as hardware \cite{ThomasEtAl:2012}. Thus it is widely agreed that the most crucial factor for software quality is the underlying development process. In particular, a posteriori testing is not sufficient for quality assurance.

Bypassing the entire ISO 26262 development process altogether is permitted only in exceptional cases. A common possibility is to establish a \emph{proven-in-use} argument \cite[part 8.14]{ISO26262:2011}. The proven-in-use candidate must have been field-tested, previously used in other vehicle models or in other safety-related industries, or it must be a component off-the-shelf (COTS) that is not necessarily intended for automotive applications. The argument must show that the observed incident rate is below the prescribed level for the corresponding ASIL (ASIL D: 1 fit, ASIL C: 10 fit, ASIL B: 100 fit, ASIL A: 1,000 fit).\\

\noindent\fbox{\begin{minipage}{\dimexpr\textwidth-2\fboxsep-3\fboxrule\relax}
\textbf{(G) V-model:} The safety process of ISO 26262 is based on a safety V-model. As such it is not straightforward to match with an agile development process, which is the natural choice for AV development.
\end{minipage}}


\subsection{Freedom from Interference}

An important aspect of hardware and software development is \emph{freedom from interference} between software components (SWCs) \cite[part 6-D]{ISO26262:2011}. If multiple SWCs are placed on the same ECU, the entire software on this ECU must be developed according to the highest ASIL. This generally leads to an \emph{ASIL lift-up} of SWCs with a lower ASIL, unless the SWCs are placed on different partitions for which \emph{freedom from interference} can be ensured: 
\vspace*{-0.1cm}
\begin{itemize}
	\setlength\itemsep{1pt}
	\item The memory of each partition is protected from read/write access from another partition, by a \emph{memory protection unit}.
	\item The dedicated execution order and execution time for each task is preserved by a central \emph{watchdog}.
	\item The communication between SWCs on different ECUs must be secured by an \emph{end-to-end protection}, to exclude interferences by any intermediary software.
\end{itemize}

\vspace*{0.1cm}
\noindent\fbox{\begin{minipage}{\dimexpr\textwidth-2\fboxsep-3\fboxrule\relax}
\textbf{(H) Centralized architecture:} AVs typically have a centralized architecture. Establishing freedom from interference between SWCs can represent a big challenge. 
\end{minipage}}

\subsection{Tool Qualification and Software Libraries}

Besides the hardware and software components themselves, the software tools used in their development may have to be qualified \cite[part 8.11]{ISO26262:2011}. In particular, it must be evaluated to what extent a tool may introduce errors that compromise the safety of the item. As ISO 26262 is not concerned with license issues, it does allow the integration of external (e.g., open source) software modules or libraries \cite[part 8.12]{ISO26262:2011}, if they have been developed according to the necessary safety standards \cite{Kriso:2012}.\\

\noindent\fbox{\begin{minipage}{\dimexpr\textwidth-2\fboxsep-3\fboxrule\relax}
\textbf{(I) Tools and libraries:} Software development for AVs often relies heavily on open source tools and libraries that have not been developed in accordance with the requirements of ISO 26262.
\end{minipage}}

\section{DISCUSSION}\label{Sec:Discussion}

This paper has outlined the safety process of ISO 26262 for the purpose of control design. It has shown some of the potential difficulties with its application to AVs. The main question that is naturally raised is about the validity of the analytic approach of ISO 26262 to FS. The approach leverages decades of experiences, but it now faces a paradigm shift with the emergence of AV technology.

In the view of the author, some of the issues mentioned in this paper have been recognized, with possible remedies under development. In particular, the topic of SOTIF (A) is addressed by a new publicly available standard (PAS) \cite{SOTIF:2017} that will complement the second edition of ISO 26262. The question of security (A) is largely orthogonal to FS and SOTIF and can therefore be tackled independently. Some of the difficulties with the item definition (B) can be handled by restricting the AV to specific, well-defined environments \cite{KoopWag:2016}. The fail-operational condition (E) will be addressed by a second edition of ISO 26262, due to be published in 2018, where \emph{availability} is considered specifically as a property of a system. This second edition of ISO 26262 also features a more abstract and objective safety assessment, in order to simplify its application, e.g., in an agile development process (G).

For some of the above challenges, however, new concepts are needed. For instance, safety validation (C) requires new methods for empirical testing, based on simulation and prototypes. However, a brute-force statistical approach will not work \cite{KoopWag:2016}, as the amount of testing effort becomes enormous. In a strict sense, no changes to the system can be made during these tests. Moreover, the failure rates stated in ISO 26262 are generally \emph{not} sufficiently low for the acceptance of AVs. In fact, they represent target goals for the development process, yet they are eventually far exceeded by the actual system developed under ISO 26262. Finally, it will be necessary to identify the root causes of safety issues, from an ethical and legal point of view. 

The remaining issues (D,F,H,I) are, by and large, related to the general complexity of AVs. To this end, some general strategies for handling complexity may be applicable. First, the design of a sufficiently modular architecture with well-defined components and interfaces may help with partly decentralizing the safety validation process. Second, with the (components') technology maturing, some of the technical issues and their impact on safety will be better understood. Third, new development tools and software libraries are needed that support the abstraction of low-level development issues. These tools also need to gain maturity and their safety has to be certified.
\vspace*{0.5cm}

\textbf{Acknowledgements.} The author would like to thank David Ward, Marc Blumentritt, and four anonymous reviewers for helpful comments and discussions.


\bibliographystyle{IEEEtran}
\bibliography{bibmath,bibcontr,bibeng}


\end{document}